\pdfoutput=1
\newcommand*{\ATLASLATEXPATH}{latex/}
\documentclass[cernpreprint, atlasdraft=false, texlive=2016, american,coverpage=no]{\ATLASLATEXPATH atlasdoc}

\usepackage[subfigure]{\ATLASLATEXPATH atlaspackage }
\usepackage{\ATLASLATEXPATH atlasbiblatex}

\usepackage{\ATLASLATEXPATH atlasphysics}

\graphicspath{{logos/}{figures/}}

\usepackage{atlas-document-defs}

\AtlasTitle{Search for resonances in the mass distribution of jet pairs with one or two jets identified as $\boldmath b$-jets in proton--proton collisions at $\sqrt{s}=13~\TeV$ with the ATLAS detector}

\AtlasAbstract{%
A search for new resonances decaying into jets containing $b$-hadrons in $pp$ collisions with the ATLAS detector at the LHC is presented in the dijet mass range from $0.57~\TeV$ to $7~\TeV$.  The dataset corresponds to an integrated luminosity of up to 36.1~fb$^{-1}$ collected in 2015 and 2016 at $\sqrt{s} = 13~\TeV$.  No evidence of a significant excess of events above the smooth background shape is found.  Upper cross-section limits and lower limits on the corresponding signal mass parameters for several types of signal hypotheses are provided at 95\% CL.  In addition, 95\% CL upper limits are set on the cross-sections for new processes that would produce Gaussian-shaped signals in the di-$b$-jet mass distributions.
}

\AtlasRefCode{EXOT-2016-33}

\PreprintIdNumber{CERN-EP-2018-075}

\HepDataRecord{83179}

\AtlasJournal{\PRD}
\AtlasJournalRef{\PRD 98, 032016 (2018)}
\AtlasDOI{10.1103/PhysRevD.98.032016}

\hypersetup{pdftitle={ATLAS document},pdfauthor={The ATLAS Collaboration}}

\begin{document}

\maketitle


\section{Introduction}
\label{sec:intro}

New heavy particles that couple to quarks or gluons are predicted by several extensions of the Standard Model (SM)~\cite{Zprime1,Zprime2,Zprime3,qstar1,qstar2}.
There is a renewed interest as these new particles can act as mediators for dark matter (DM) interactions~\cite{DM1,DM2,DM3,DM4,DM5}.
Such heavy particles can be produced in proton-proton collisions at relatively high rates thanks to their possibly strong coupling. The new particles can decay into quarks and gluons, that hadronize and form jets that are observable in the detector. Such a decay will produce dijet systems with an invariant mass around the mass of the new particle, appearing as an excess above the continuum background.
This analysis searches for a resonant excess in the dijet mass distribution.

The dijet mass range explored in previous analyses depends on the available center-of-mass energy as well as on the size of the data sample. 
Past dijet searches have investigated the dijet mass ranges 110--350~\GeV\ at the $\textrm Sp\bar{p}S$ collider~\cite{dijetUA1} and 260--1400~\GeV~\cite{CDF:2009DijetSearch}, 250--1100~\GeV~\cite{DZero:2009DijetAng} at the Tevatron. At the LHC, the most recent CMS search covers 0.6--7.5~\TeV~\cite{Sirunyan:2016iap}, while the last ATLAS search covers 0.45--6.5~\TeV~\cite{Aaboud:2017yvp,Aaboud:2018fzt}.

Searches restricted to final states involving jets identified as containing a $b$-hadron have an increased sensitivity to certain scenarios, for example to particles that preferentially decay into $b\bar{b}$ quark pairs as predicted by some dark-matter models~\cite{Fermi, Fermi2}.
But the sensitivity can be improved even for resonances without an enhanced $b\bar{b}$ decay mode, like many $Z'$ models described below, if the search suffers from non-$q\bar{q}$ backgrounds, in particular gluon radiation.
Such searches have been performed by CDF covering the mass range 250--750~\GeV~\cite{Abe:1998uz}, by CMS covering 0.3--4~\TeV~\cite{CMS:2012yf,Sirunyan:2018pas} and by ATLAS  covering 1--5~\TeV~\cite{EXOT-2015-22}. So far no deviations from the Standard Model have been found.

Compared to previous collider searches that have explored the mass region below 1~\TeV, the LHC can provide higher sensitivity and cover yet unexplored coupling values due to the increase in parton luminosity~\cite{ATLASisr}. Consequently resonance searches in this mass range are still of interest. In particular, some dark-matter models predict such particles~\cite{Fermi, Fermi2}.  In this paper an extension of the ATLAS search into this lower-mass region is made possible by a new trigger strategy, identifying two $b$-jets at trigger level. This strategy is able to cope with the large event rate in the lower dijet-mass region.  The search presented in this paper probes the mass range 0.57--7~\TeV.

\begin{figure}[ht]
\begin{center}
   \subfigure[] {\includegraphics[scale=0.4, angle=0]{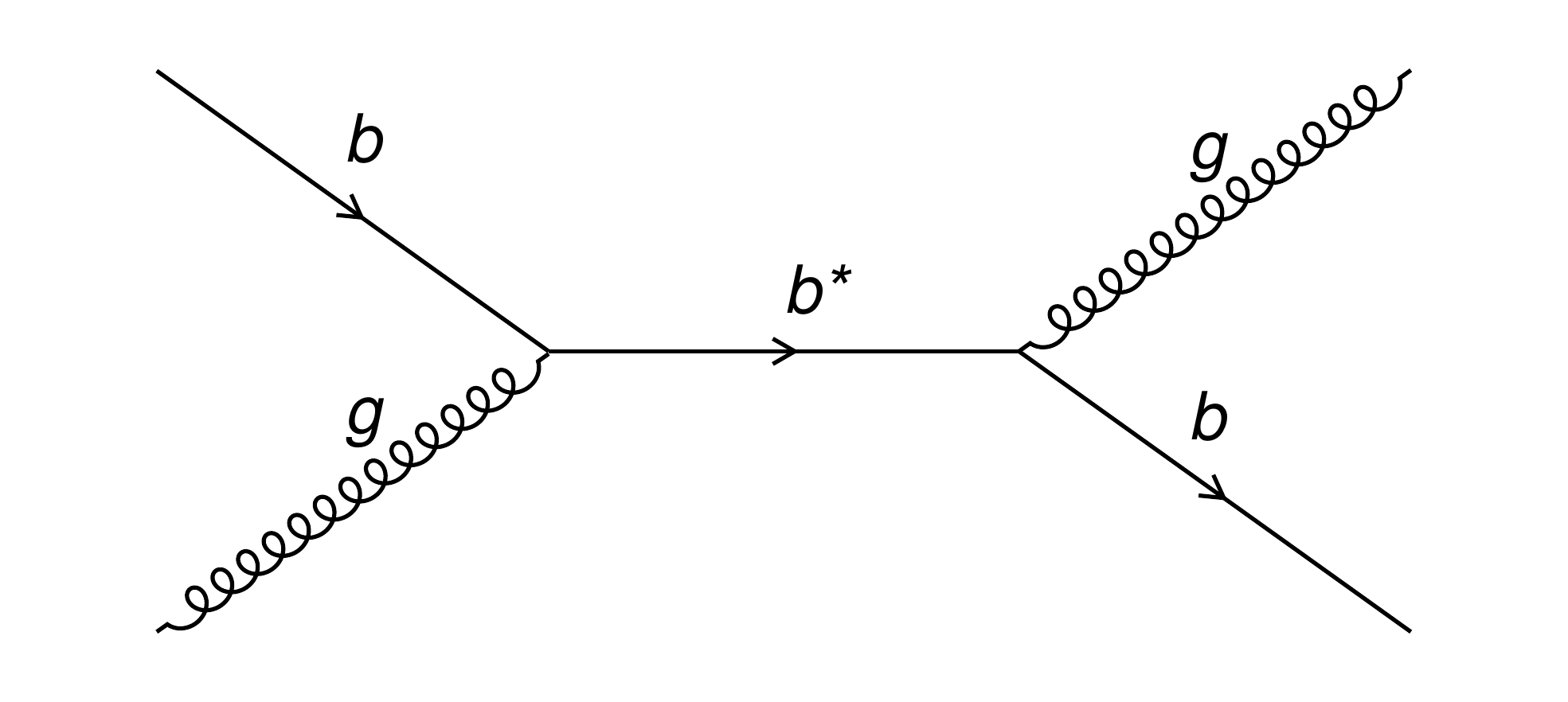}\hfill}\subfigure[]{
   \includegraphics[scale=0.4, angle=0]{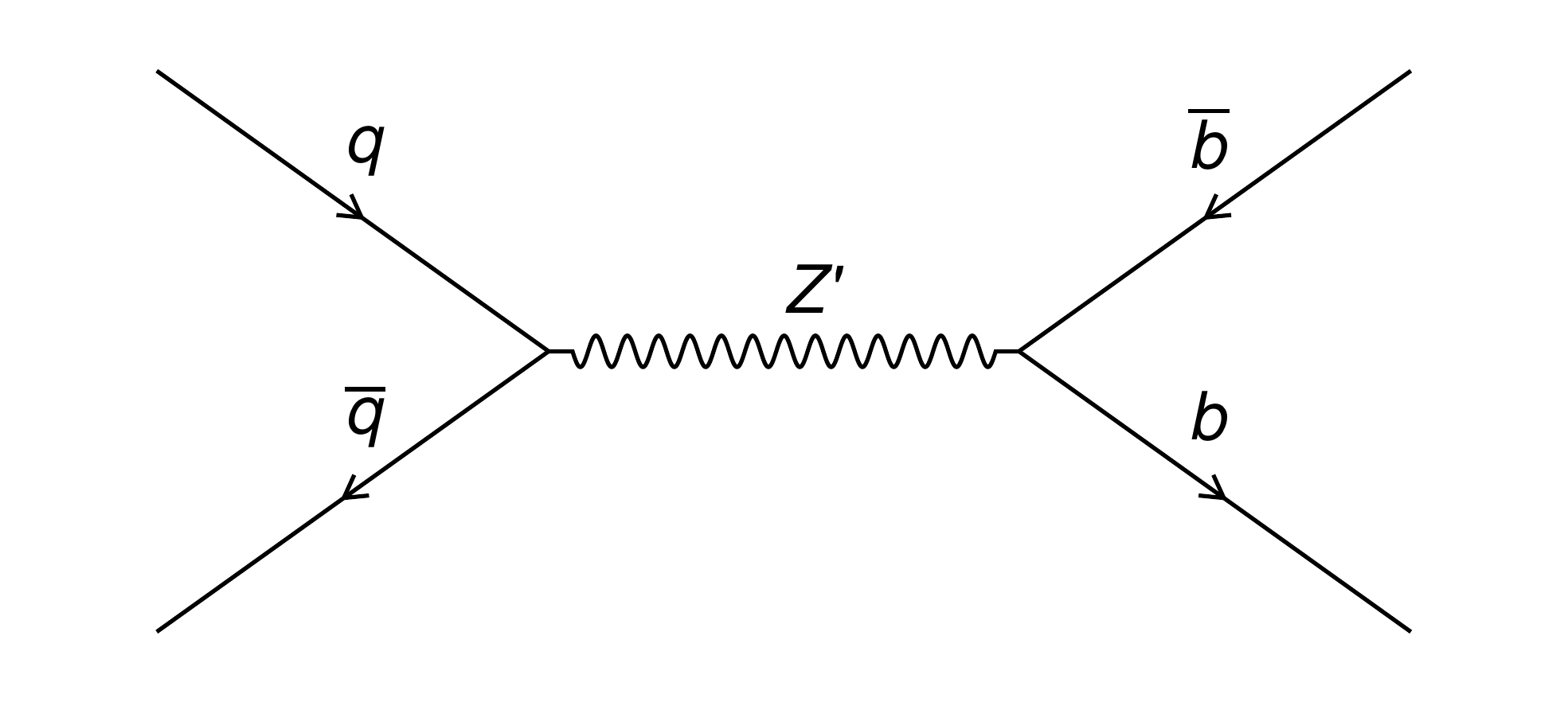}}
\end{center}
\caption{Example of the leading-order Feynman diagram for production and decay of (a) $b^*$  and (b) $Z'$ into final states involving $b$ quarks.}
\label{fig:FeynmanPlot}
\end{figure}

The results are interpreted in the context of several benchmark models.
An excited $b^*$-quark, with a dominant decay mode to $bg$, is used as the benchmark for events with at least one jet identified containing $b$-hadrons: the $\ge 1$ $b$-tag category. Excited quarks arise from compositeness models~\cite{qstar1,qstar2}.
Models featuring an additional gauge boson called $Z'$~\cite{Zprime1,Zprime2,Zprime3}, including a dark-matter model with a $Z'$ mediator~\cite{DM1,DM2}, are considered in the two $b$-tags category. The leading-order Feynman diagrams for these processes are shown in Fig.~\ref{fig:FeynmanPlot}. Further details can be found in Sec.~\ref{sec:mc}. In addition, model-independent limits are set on generic resonance signals that have a Gaussian reconstructed shape.
These limits assume, after applying the selection, a narrow-resonance signal shape with an intrinsic width that can be safely truncated or neglected, so that the reconstructed mass distribution reflects the experimental resolution and can be approximated by a Gaussian distribution ~\cite{EXOT-2013-11}.

\section{ATLAS detector}
\label{sec:detector}

\newcommand{\AtlasCoordFootnote}{%
ATLAS uses a right-handed coordinate system with its origin at the nominal interaction point (IP)
in the center of the detector and the $z$ axis along the beam pipe.
The $x$ axis points from the IP to the center of the LHC ring,
and the $y$ axis points upwards.
Cylindrical coordinates $(r,\phi)$ are used in the transverse plane, 
$\phi$ being the azimuthal angle around the $z$ axis.
The pseudorapidity is defined in terms of the polar angle $\theta$ as $\eta = -\ln \tan(\theta/2)$.
Angular distance is measured in units of $\Delta R \equiv \sqrt{(\Delta\eta)^{2} + (\Delta\phi)^{2}}$.}

\label{sec:atlas2}

The ATLAS detector~\cite{PERF-2007-01} at the LHC covers nearly the entire solid angle around the collision point.\footnote{ATLAS uses a right-handed coordinate system with its origin at the nominal interaction point (IP) in the center of the detector and the $z$ axis along the beam pipe. The $x$ axis points from the IP to the center of the LHC ring, and the $y$ axis points upward. Cylindrical coordinates $(r,\phi)$ are used in the transverse plane, $\phi$ being the azimuthal angle around the $z$ axis. The pseudorapidity is defined in terms of the polar angle $\theta$ as $\eta=-\ln\tan(\theta/2)$. Angular distance is measured in units of $\Delta R \equiv \sqrt{(\Delta\eta)^{2} + (\Delta\phi)^{2}}$.} 
It consists of an inner tracking detector surrounded by a thin superconducting solenoid, electromagnetic and hadronic calorimeters,
and a muon spectrometer incorporating three large superconducting toroid magnets.
The inner-detector system (ID) is immersed in a \SI{2}{\tesla} axial magnetic field 
and provides charged-particle tracking in the range $|\eta| < 2.5$.

A high-granularity silicon pixel detector covers the vertex region and typically provides three measurements per track.
A new inner pixel layer, the insertable B-layer~\cite{ibl,iblAdd}, was added during the 2013-2014 LHC shutdown.
It is located at a mean sensor radius of 32~mm from the beam line,
providing a fourth pixel hit.
The pixel detector is followed by a silicon microstrip tracker, which usually provides four two-dimensional measurement points per track.
These silicon detectors are complemented by a transition radiation tracker,
which enables radially extended track reconstruction up to $|\eta| = 2.0$. 
The transition radiation tracker also provides electron identification information 
based on the fraction of hits (typically 30 in total) that deposit energy above a threshold corresponding to transition radiation.

The calorimeter system covers the pseudorapidity range $|\eta| < 4.9$.
Within the region $|\eta|< 3.2$, electromagnetic calorimetry is provided by barrel and 
endcap high-granularity lead/liquid-argon (LAr) electromagnetic calorimeters,
with an additional thin LAr presampler covering $|\eta| < 1.8$,
to correct for energy loss in material upstream of the calorimeters.
Hadronic calorimetry is provided by the steel/scintillator-tile calorimeter,
segmented into three barrel structures within $|\eta| < 1.7$, and two copper/LAr hadronic endcap calorimeters.
The solid angle coverage is completed with forward copper/LAr and tungsten/LAr calorimeter modules
optimized for electromagnetic and hadronic measurements respectively.

A two-level trigger system is used to select interesting events.
The first trigger level is implemented in hardware and uses a subset of detector information
to reduce the event rate to a design value of at most \SI{100}{\kHz}.
This is followed by a software-based high-level trigger (HLT) which reduces the event rate to about \SI{1}{\kHz}.

\section{Simulated signal samples}
\label{sec:mc}

The Monte Carlo (MC) simulation is used to generate samples describing the benchmark signal models under consideration.
These signal samples were generated with {\PYTHIA}8~\cite{pythia8} using the A14 set of tuned parameters~\cite{a14tune} and the NNPDF2.3 PDF set~\cite{nnpdf}. The \textsc{EvtGen} decay package~\cite{evtgen} is used for bottom and charm hadron decays.
The generated samples were processed with the ATLAS detector simulation~\cite{atlassim}, which is based on the {\GEANT}4 package~\cite{geant4}. 
To account for additional proton-proton interactions (pileup), further minimum-bias interactions were generated using {\PYTHIA}8 and the MSTW2008LO PDF set~\cite{mstw} and superimposed on the hard-scattering events. The MC samples were reweighted to match the distribution of the number of collisions per bunch crossing observed in the data. For basic background validation a leading-order multijet sample was generated with {\PYTHIA}8 and the same parameters and PDF set used for the signal models. The same reconstruction software was run on the simulated events as was used for recorded collision data.

Signal events in the excited $b^*$-quark model were generated with the compositeness scale $\Lambda$ set to the excited-quark mass $m_{b^*}$ and
an intrinsic decay width of $\Gamma\sim 0.006\times m_{b^*}$. The branching fraction for the dominant decay $b^*\rightarrow bg$ is 85\%. 
The remaining decay modes are $b^*\rightarrow b\gamma$, $b^*\rightarrow bZ^0$ and $b^*\rightarrow tW^-$. 
The leading-order (LO) cross section for a $2.5~\TeV$ $b^*$-quark is 123~fb~\cite{pythia8}.

Three models with a $Z'$ gauge boson are considered.
In the sequential standard model (SSM), the $Z'$ boson has the same couplings to SM fermions as the SM $Z$ boson and the bottom-quark decay branching fraction ${\mathcal{B}}(Z'\rightarrow b\bar{b})$ is $13.8\%$. The leptophobic $Z'$ model differs by having vanishing couplings to leptons. The corresponding value of ${\mathcal{B}}(Z'\rightarrow b\bar{b})$ is $18.9\%$ in the mass ranges considered.  The intrinsic width of the $Z'$ bosons are set to $3\%$ of the resonance mass~\cite{Zprime1}.
The leading-order {\PYTHIA}8 SSM and leptophobic $Z'$ cross sections were corrected to next-to-leading order (NLO) using cross-sections calculated at LO and at NLO using {\MADGRAPH}5\_aMC@NLO~\cite{madgraph}, with the NNPDF2.3 LO and NLO PDF sets, respectively. The NLO prediction uses a model of neutral vector bosons implemented in FeynRules~\cite{FeynRules} with NLO terms evaluated via NLOCT package~\cite{NLOCT}. The NLO cross section times branching fraction ${\sigma\cdot\mathcal{B}}(Z'\rightarrow b\bar{b})$ for a $2~\TeV$ SSM neutral vector boson is $0.10$~fb~\cite{pythia8,madgraph}. For both models, only decays into $b$-quark pairs were simulated.
 
Lastly, a simplified dark-matter model~\cite{DM4} with a $Z'$ axial-vector mediator is considered. The mediator to SM quark coupling ($g_\text{SM}$) was set to 0.1 or 0.25, the mediator to axial DM coupling to 1.0 and the mass of the dark-matter particle was fixed to $10~\TeV$ within the scope of the reference~\cite{DM4}. The intrinsic width was calculated by {\MADGRAPH}5\_aMC@NLO~\cite{madgraph}. The LO cross section times branching fraction ${\sigma\cdot\mathcal{B}}(Z'\rightarrow q\bar{q})$ for a $1~\TeV$  axial-vector mediator with $g_\text{SM}=0.1$ is $2.7$~fb~\cite{pythia8}.


\section{Data samples and event selection}
\label{sec:sample}

The data for this analysis were collected by the ATLAS detector in $pp$ collisions with a center-of-mass energy of $\sqrt{s} = 13~\TeV$. The data set for the high dijet-mass region $m_{jj} > 1.2~\TeV$ was recorded by selecting events from an inclusive jet trigger requiring at least one jet with a transverse momentum $\pt$ above $380~\GeV$, and corresponds to an integrated luminosity of $3.2~\ifb$ in 2015 and $\lumi~\ifb$ in 2016. Events for the low dijet-mass region $570~\GeV < m_{jj} < 1.5~\TeV$ were recorded using a dijet trigger employing an online algorithm to identify two jets containing $b$-hadrons and having transverse momentum $\pt$ above $150~\GeV$ and $50~\GeV$, respectively. The above transverse momentum requirements are fully efficient in the quoted mass range. This trigger overcomes the limitation related to the high inclusive single jet trigger rate. Because the $b$-jet trigger was active only for parts of the data taking period, the total integrated luminosity that the low dijet-mass sample corresponds to is $\lumilow~\ifb$ in 2016. The $b$-jet trigger chain~\cite{TRIG-2016-01} starts by requiring an energy deposit measured with coarse granularity ($\Delta\phi\times\Delta\eta=0.2\times 0.2$) in the calorimeter at the first trigger level.  In the HLT, a two-step tracking algorithm is run. First, a fast tracking stage is used to find the primary vertex of the event. The results from this first stage seed precision tracking. The output of this tracking stage provides the input for the $b$-jet identification algorithms, which are based on the offline tools described further below. The identification efficiency is 60\% per $b$-jet at trigger level when integrated over transverse momentum $\pT$ and pseudorapidity $\eta$.

Offline jets are reconstructed from topological clusters of energy deposits in the calorimeters~\cite{PERF-2014-07} with the anti-$k_t$ algorithm~\cite{antikt, Fastjet} with a radius parameter of 0.4. Jet energies and directions are corrected by the jet calibrations as described in Ref.~\cite{ATL-PHYS-PUB-2015-015}. Jets containing a $b$-hadron are identified using a multivariate algorithm~\cite{PERF-2012-04, ATL-PHYS-PUB-2016-012}. This algorithm makes use of the impact parameters of tracks and the reconstructed displaced vertices in the ID.  The offline $b$-tagging efficiency operating points are determined on a $t\bar{t}$ sample when integrated over $\pT$ and $\eta$ \cite{Aaboud:2018xwy}. 
In the high-mass region, an $85\%$ efficiency offline $b$-tagging operating point is employed. In the low-mass region, a  $70\%$ offline efficiency  $b$-tagging operating point is adopted in addition to the online $b$-tagging requirement, because the online $b$-identification is only partially correlated to the offline $b$-tagging. The online $b$-tagging algorithm is not fully emulated in MC and the tagging efficiency is needed to estimate the signal acceptance. The online $b$-tagging efficiency is measured using a high $b$-jet purity dilepton $t\bar{t}$ sample.
The offline $b$-tagging operating points have been optimized in order to maximize the overall sensitivity.


\begin{figure}[!ht]
\begin{center}
\subfigure[High-mass region]{
\includegraphics[width=0.48\linewidth, angle=0]{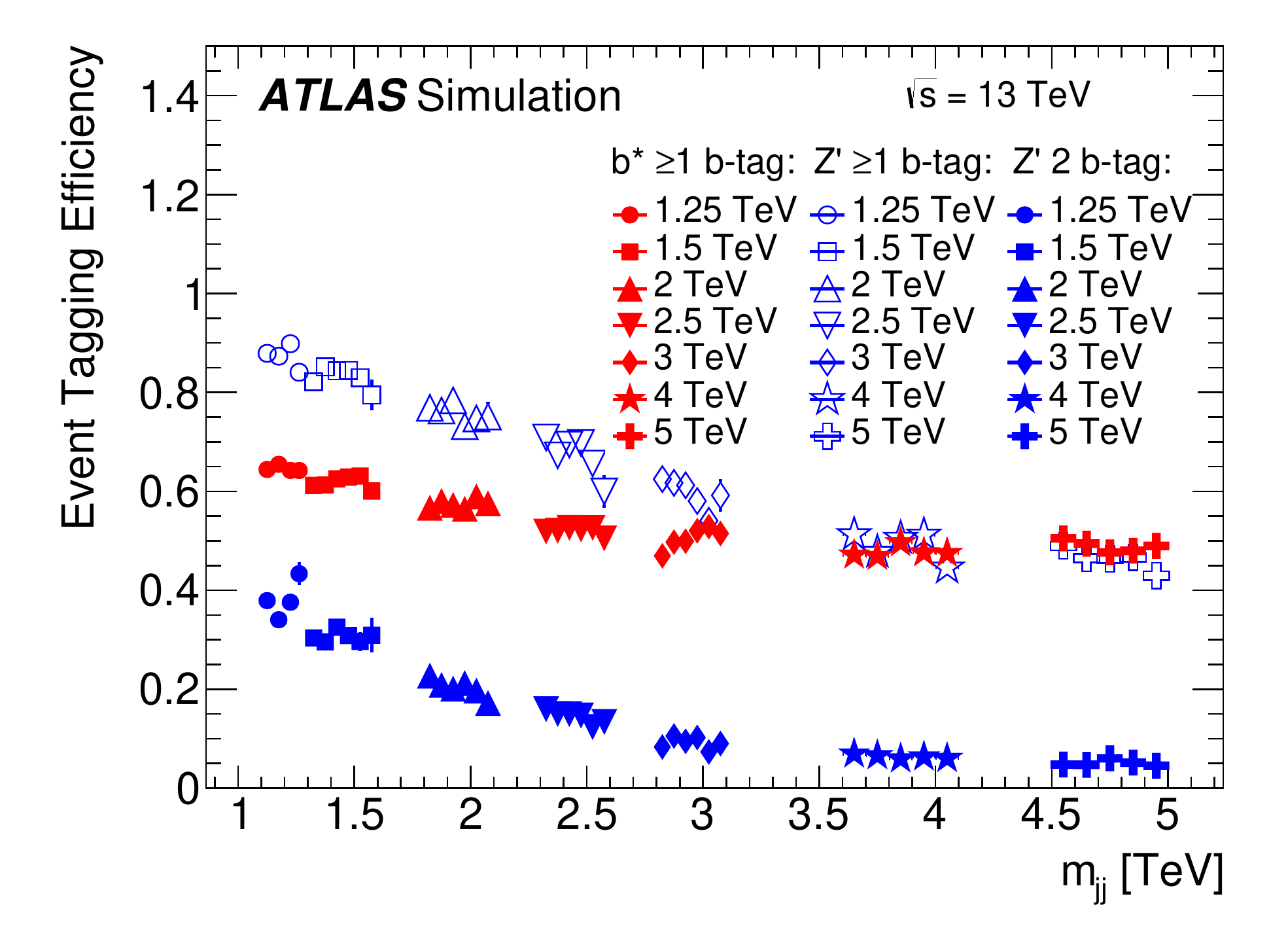}}
\subfigure[Low-mass region]{
\includegraphics[width=0.48\linewidth, angle=0]{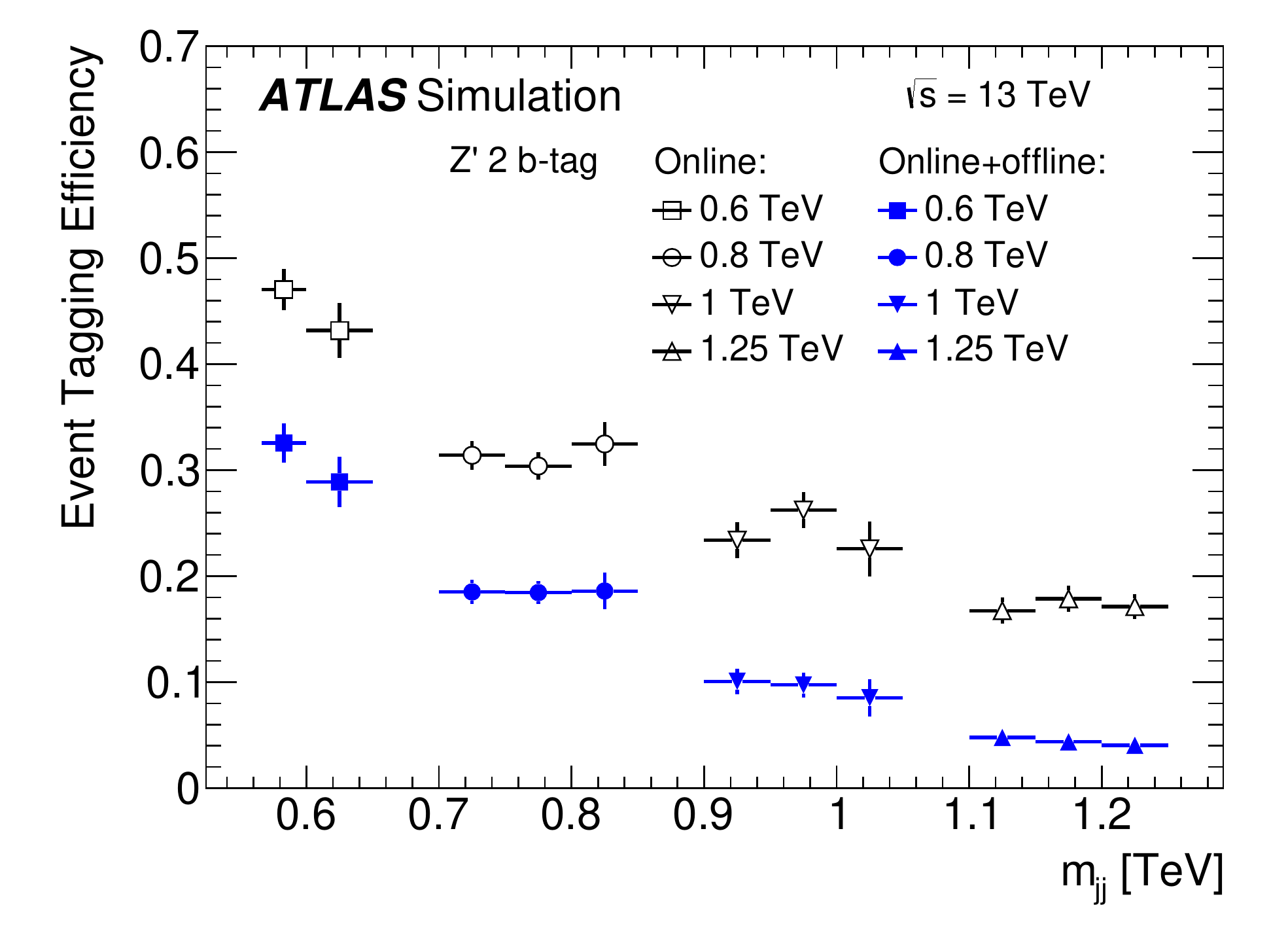}}
\end{center}
\caption{The per-event $b$-tagging efficiencies after the event selection, as a function of the reconstructed invariant mass, $m_{jj}$.
Events are classified into single $b$-tagged ``1b'' or double $b$-tagged ``2b'' categories.  The efficiencies are shown for simulated event samples corresponding to (a) seven different $b^*$ and $Z'$ resonance masses in the high-mass region and (b) four different $Z'$ resonance masses in the low-mass region. For each generated resonance mass, several reconstructed mass points around the resonance mass are shown. Efficiencies of identifying an event with two $b$-jets at trigger level only (online) and when requiring offline confirmation (online+offline) are shown.}
\label{fig:eventtageff}
\end{figure}

In order to ensure full trigger efficiency and lower pileup contamination, the event selection requires  a minimum transverse momentum of $\pT > 430~\GeV$ and $\pT > 80~\GeV$ for the leading and subleading jet, respectively. The requirement on the leading jet is relaxed to $200~\GeV$ for the low-mass region, corresponding to the reduced transverse momentum requirement in the trigger. Both jets are required to have pseudorapidity $|\eta|<2.0$ to allow fully efficient $b$-jet identification in the two mass regions. To reduce background from multijet production and enhance s-channel signal processes, the rapidity difference $y^*=(y_1-y_2)/2$ between the two leading jets is required to be $|y^*|<0.8$. In the low-mass region this requirement is tightened to $|y^*|<0.6$ to avoid regions of reduced trigger efficiency at the lower mass boundary.

In the analysis, one or both of the leading jets are required to be identified as $b$-jets.
The per-event efficiencies, taking the $b$-tagging requirement(s) into account, are shown as functions of the reconstructed invariant mass of the two leading jets, $m_{jj}$, for several signal models in Fig.~\ref{fig:eventtageff}. 
Events from the $Z'$ model have a higher event-tagging efficiency than for $b^*$ events in the inclusive ``1b'' category because  $Z'$ events contain two $b$-quarks in the final state. In the high-mass region, the $b^*\to bg$ decay can be followed by the gluon splitting into a $b\bar{b}$ pair, which therefore enhances the event $b$-tagging efficiency for $b^*$ events relative to the $Z'$ signal.

\section{Analysis}
\label{sec:analysis}
%

The observed dijet mass distribution of the two leading jets in the high-mass event selection ($m_{jj} > 1.2~\TeV$), where at least one ($\ge 1$ $b$-tag) or both (2 $b$-tags) jets are identified as $b$-jets, is inspected for resonant contributions from new-physics scenarios.  In the low-mass analysis ($570~\GeV < m_{jj} < 1.5~\TeV$) only a selection with two $b$-tags is considered due to the trigger selection. The treatment of the 2-$b$-tags overlap region ($1.2~\TeV < m_{jj} < 1.5~\TeV$) is discussed in Sec.~\ref{sec:result}.

The dominant background arises from multijet final states.  While the shape of the $m_{jj}$ distribution in data is found to be in good agreement with the {\PYTHIA}8 multijet MC simulation, the normalization is not.  In this analysis the background is evaluated from a fit to the mass distribution in data.  

Previous dijet resonance searches~\cite{EXOT-2015-02, Aaboud:2017yvp} have found that the following fit function:
\begin{equation}
f(x)=p_1(1-x)^{p_2}(x)^{p_3+p_4\ln{x}+p_5(\ln{x})^{2}},
\label{fit}
\end{equation}
where $p_i$ are fit parameters and $x\equiv m_{jj}/\sqrt{s}$, provides a good global fit to dijet mass distributions in data as well as leading-order and next-to-leading-order simulations of QCD dijet production, where $p_5\equiv p_4\equiv 0$~\cite{EXOT-2015-02} or $p_5\equiv 0$~\cite{Aaboud:2017yvp}.  However, it is found that Eq.~(\ref{fit}) no longer provides an adequate description of the data for the whole mass distribution comprising the high-mass and low-mass regions.  This effect is attributed to a larger data sample than in previous analyses that employed the global fit strategy, in conjunction with the shaping of the $b$-tagged dijet mass distribution due to the \pT dependence of the $b$-tagging efficiency and variations of the quark flavor fractions as a function of \pT.
The background estimate is therefore derived from a sliding-window fit by using the fit function from Eq.~(\ref{fit}) with four or five fit parameters, and by fitting only restricted regions of the spectrum at a time.  This technique was introduced in the most recent ATLAS dijet resonance search~\cite{Aaboud:2017yvp} and is briefly described here.  The number of fit parameters of the sliding-window fit are chosen to have the largest possible window size for a fit function with the fewest number of parameters.
The four-parameter fit [where $p_5$ is set to zero in Eq.~(\ref{fit})] is chosen for the high-mass 2-$b$-tags selection, while the five-parameter fit is chosen for the low-mass and the inclusive $\ge 1$ $b$-tag selections.
For the low-mass selection the window size is chosen to comprise 14 out of 31 total bins, whereas for the high-mass selection the window size corresponds to 16 bins for the 2-$b$-tags selection and 22 bins for the $\ge 1$ $b$-tag selection out of a total number of 75 bins.
The bin width follows approximately the $m_{jj}$ invariant mass resolution as derived from the MC simulation of multijet processes. The bin width increases from about $20~\GeV$ at a mass of $500~\GeV$ to about $130~\GeV$ at a mass of $7~\TeV$.
The background prediction over the full mass range is constructed in each mass bin by evaluating the fit function in the window centered around that bin.  At the low and high edge of the mass distribution, the sliding-window regions do not extend outside the considered mass range.



The validity of this background-fitting method is tested in data control regions, where no offline $b$-jet identification is required and the MC-estimated $b$-tagging efficiencies are applied as a weight .
Representative background data sets are created by injecting Poisson fluctuations into the data control regions. Spurious-signal tests are performed to verify that no artifact is created during the fitting procedure by fitting hundreds of representative background data sets, and then checking the flatness of the probability returned by the BumpHunter algorithm~\cite{bumphunter} as detailed below. 
 The fit is shown to be robust against spurious signals.
In addition, signal injection tests are performed and good linearity between the injected and extracted signal is observed for the full range of signal widths considered. No sensitivity reduction due to the choice of window size is found.



For both the low-mass and high-mass 2-$b$-tag selections the background prediction covers the full $m_{jj}$ mass region, where the lower boundaries are defined by the plateau region of the trigger as defined in Sec.~\ref{sec:sample}.  For the high-mass inclusive $\ge 1$ $b$-tag selection, studies of the validity of the fit required an increase of the lower mass boundary from $1.2$ to $1.3~\TeV$.  The largest value of $m_{jj}$ is measured to be $6.77~\TeV$ with one $b$-tag and $6.31~\TeV$ with two $b$-tags.

Fig.~\ref{fig:bhunt} shows the $m_{jj}$ distributions, overlaid with the fit results and examples of the potential signals described in Sec~\ref{sec:mc}.  The lower panel in each plot of Fig.~\ref{fig:bhunt} shows the significance of the bin-by-bin differences between the data and the fit, as calculated from Poisson probabilities, considering only statistical uncertainties. The BumpHunter algorithm is used to evaluate the statistical significance of any localized excess in the dijet mass distributions in data relative to the fitted background estimate.  The algorithm calculates the significance of any excess found in contiguous mass intervals in all possible locations of the binned $m _{jj}$ distribution, between a width of two bins and a width of half of the distribution. The intervals 3448--3749~\GeV, 3100--3235~\GeV, and 976--1068~\GeV, indicated by two vertical lines in each of the Figs.~\ref{fig:bhunt}(a)-(c), are identified as the most discrepant intervals in the inclusive 1-$b$-tag, the 2-$b$-tags high-mass, and the 2-$b$-tags low-mass region, respectively.  The purely statistical significance of each excess is evaluated using the ensemble of possible outcomes across all scanned intervals, by applying the algorithm to many pseudo-data samples drawn randomly from the background fit. The probability that statistical fluctuations of the background would produce an excess at least as significant as the one observed in the data, anywhere in the distribution, is 0.66, 0.59, and 0.57 for the inclusive 1-$b$-tag, the 2-$b$-tags high-mass, and the 2-$b$-tags low-mass region, respectively. Thus there is no evidence of a significant localized excess over the background estimate.

\begin{figure}
  \begin{center}
    \raggedright
    \subfigure[Inclusive one $b$-tag, high  mass] {
      \includegraphics[width=0.45\linewidth, angle=0]{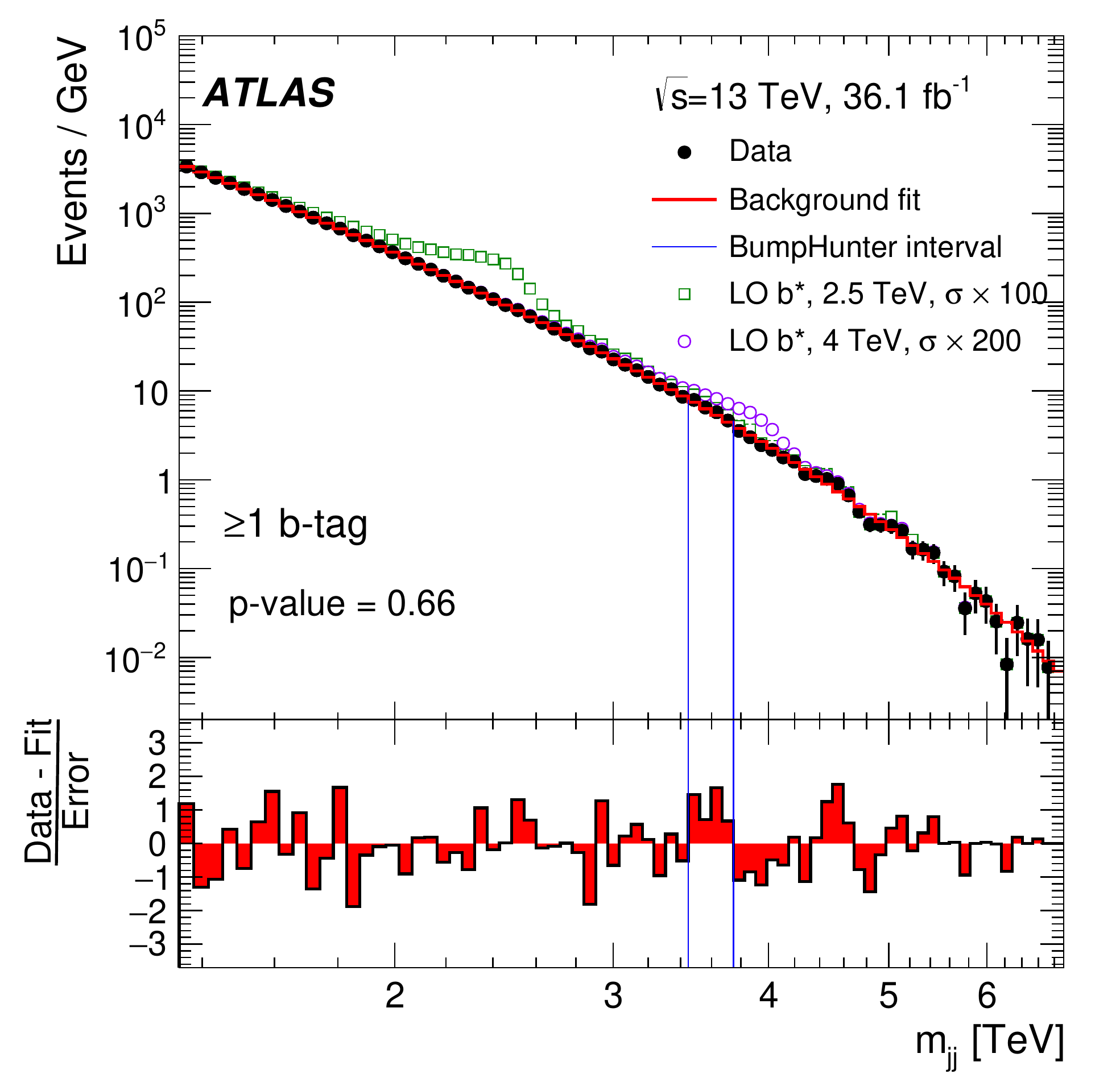}
    }\\
    \subfigure[Two $b$-tag, high  mass] {
      \includegraphics[width=0.45\linewidth, angle=0]{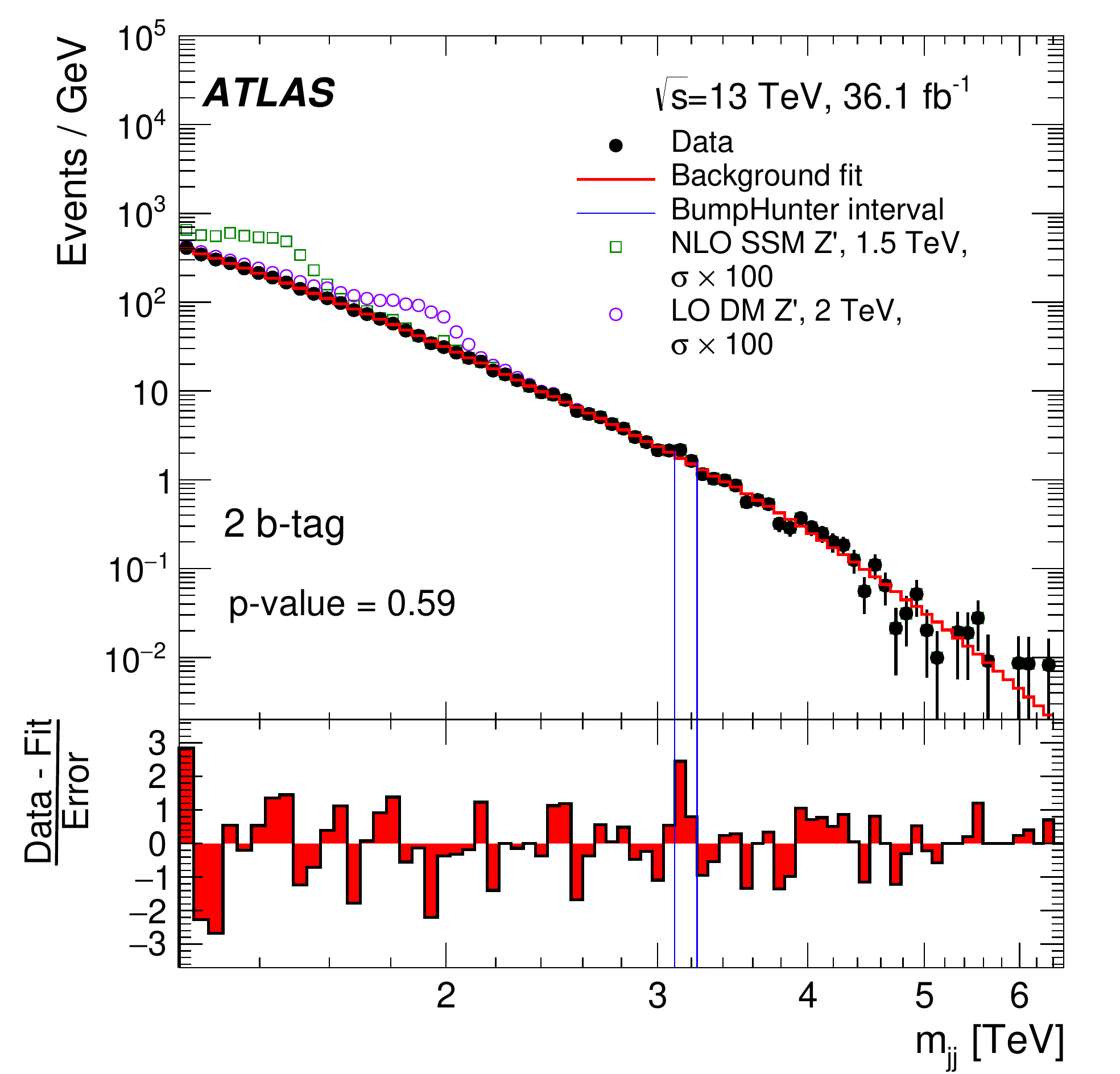}
    }
    \subfigure[Two $b$-tag, low mass] {
      \includegraphics[width=0.45\linewidth, angle=0]{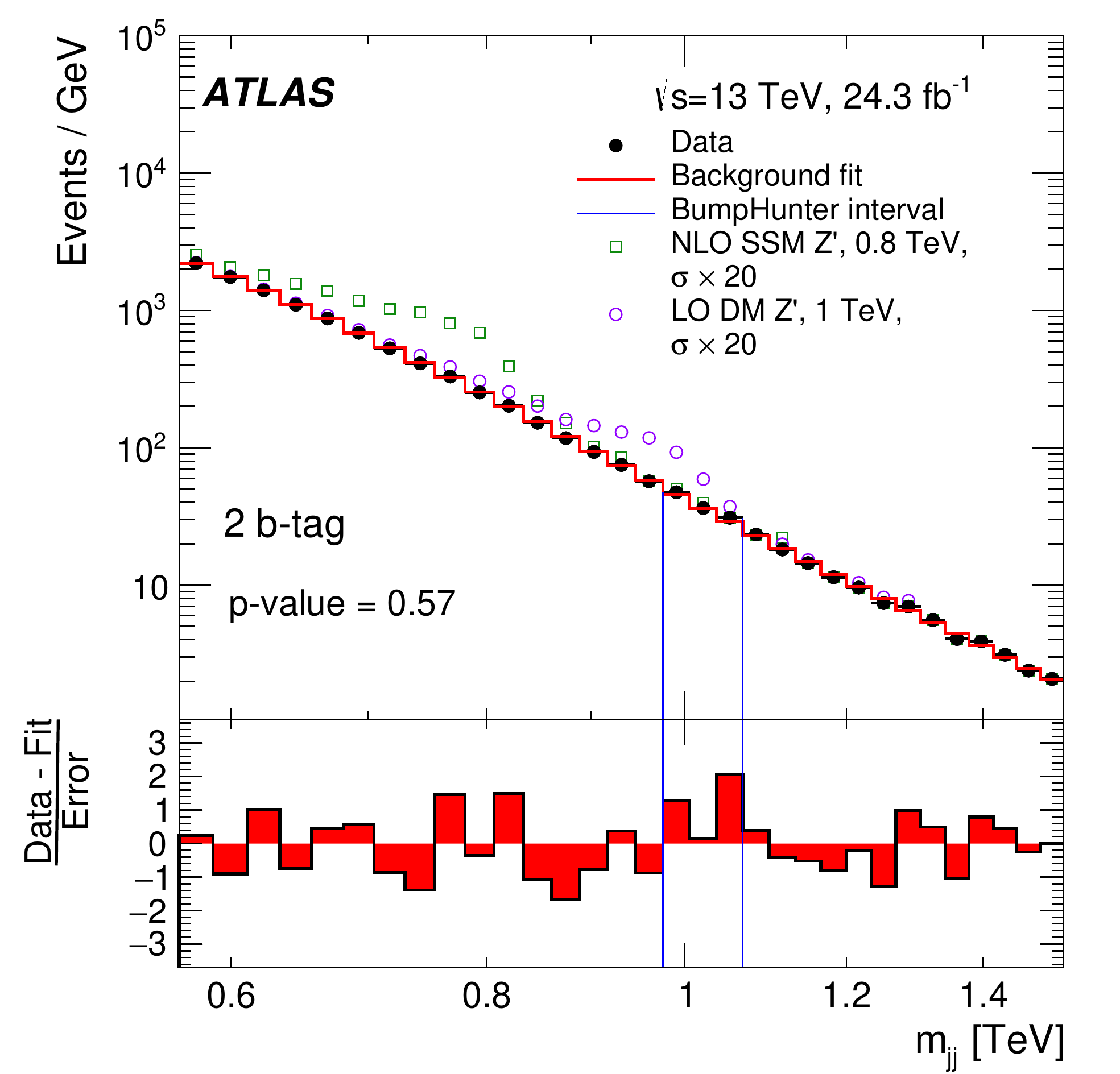}
    }
  \end{center}
\caption{Dijet mass spectra after the background only fit with the background prediction together with the result from the {\textsc BumpHunter} (see text for details). The plots show (a) the inclusive 1-$b$-tag high-mass region, (b) the high-mass region with two $b$-tags and (c) the low-mass region with two $b$-tags. The potential signals are overlaid on top of the data.
\label{fig:bhunt}}
\end{figure}

\section{Systematic uncertainties}
\label{sec:systematics}
The systematic uncertainty of the background is estimated from the uncertainty associated with the choice of the fit function and the uncertainties in the values of the fit parameters. The uncertainty due to the choice of the fit function is determined by repeating the fit procedure with one additional parameter.  For the four-parameter fit of the high-mass 2-$b$-tags selection, $p_5$ is added as an additional free parameter, and for the five-parameter fit of the low-mass and inclusive $\ge 1$ $b$-tag selections a new parameter $p_6$ is introduced in Eq.~(\ref{fit}) by redefining $x$ as $x\equiv m_{jj}/p_6$.  The uncertainty is given by the average difference between the two fit results across a set of pseudo-data drawn via Poisson fluctuations from the nominal background prediction.  The uncertainty due to the values of the fit parameters is taken to be the bin-by-bin root-mean-square of the fit results for all the pseudo-experiments using the nominal fit function. 
 
The uncertainty in the MC-based signal expectation is dominated by the uncertainty in the modeling of the $b$-tagging efficiency~\cite{PERF-2012-04,Aaboud:2018xwy}.  This uncertainty grows with jet $\pT$, with a smallest uncertainty of 2\% for jets with $\pT$ around $90~\GeV$ and up to 15\% for jet $\pT$ around $1.5~\TeV$.  The $b$-jet calibration is based on identifying a high-purity sample of $b$-jets by selecting $\ttbar$ events~\cite{Aaboud:2018xwy}.  The uncertainties are measured using data for jet $\pT<300~\GeV$ and are extrapolated to jet $\pT > 300~\GeV$ by means of MC simulation by varying quantities in the simulation that are known to affect the $b$-tagging performance, such as the track impact-parameter resolution, the fraction of poorly measured tracks, the description of the detector material, and the track multiplicity per jet.  The uncertainty in the impact-parameter resolution includes alignment effects, dead modules and additional material not properly modeled in the simulation, and is the dominant source of uncertainty for the $b$-tagging efficiency at high $\pT$.

Because the data set for the low-mass analysis is recorded using the $b$-jet trigger as described in Sec. \ref{sec:sample},  there is an additional systematic uncertainty associated with the $b$-jet trigger efficiency.  It is extracted by comparing the $b$-jet trigger efficiency in 2016 data and MC simulation in a high-purity sample of $b$-jets selected from a dilepton $\ttbar$ sample by using similar procedures to those used to measure the offline $b$-tagging efficiencies.  Uncertainties due to the mismodeling of the $b$-jet purity in simulation, mismodeling of the $b$-jet trigger efficiency for non $b$-jets, simulation statistical uncertainty, data statistical uncertainty (jet $\pT < 240~\GeV$) and simulation-based extrapolation (jet $\pT > 240~\GeV$) are taken into account. The per-jet uncertainty is estimated to be 1\%--20\% for jets with $\pT$ of 35--700~\GeV~(Fig.~\ref{fig:btageff}). 
The total uncertainty of the di-$b$-jet trigger efficiency comes from the per-jet $b$-tagging efficiency with an additional per-event uncertainty of 2\% that covers differences in the primary vertex reconstruction.

\begin{figure}[!ht]
\begin{center}
\includegraphics[width=0.5\linewidth, angle=0]{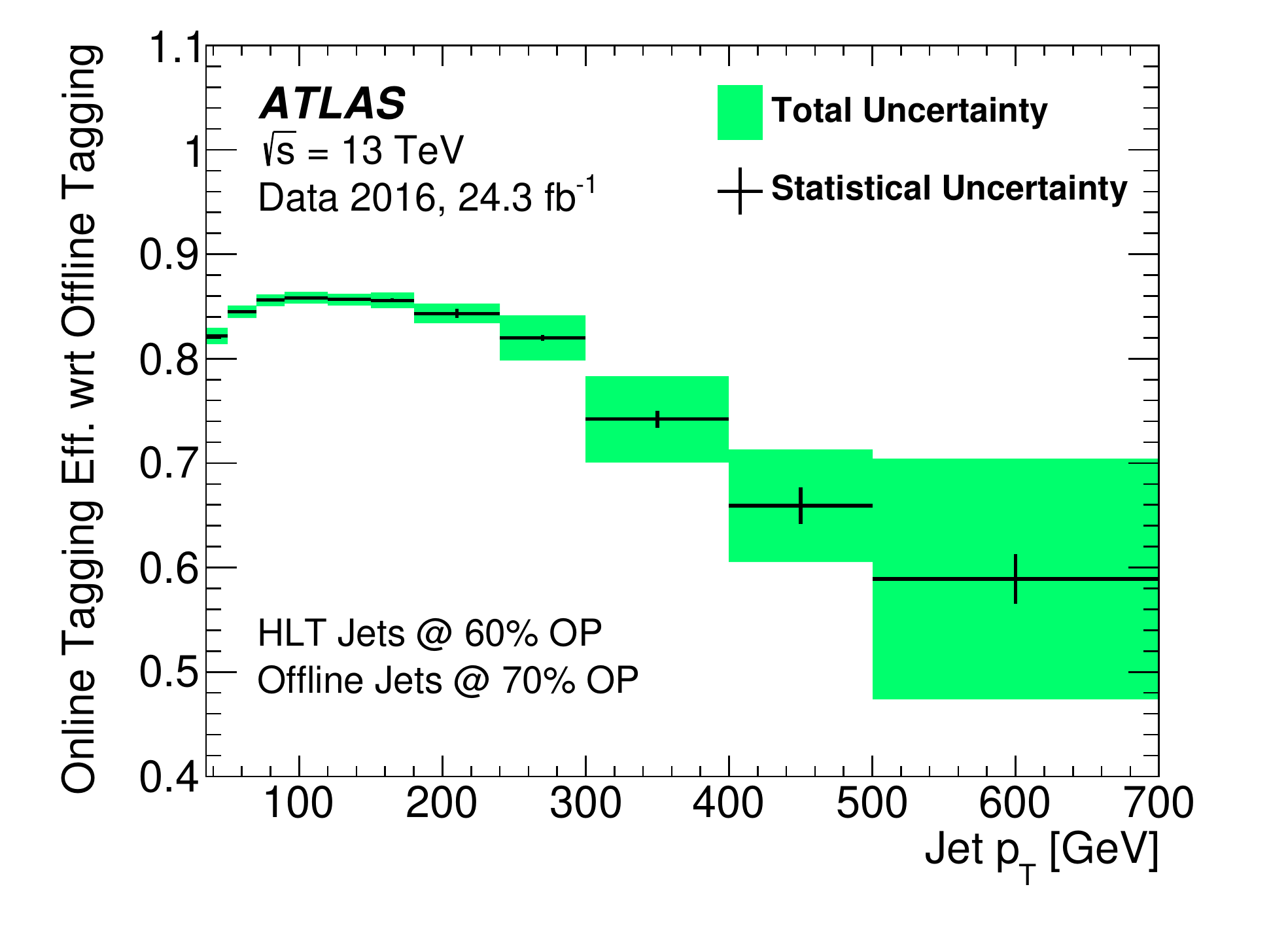}
\end{center}
\vspace{-20pt}
\caption{
The online $b$-tagging efficiency with respect to the offline b-tagging efficiency as a function of $\pT$. 
The $b$-tagging online and offline working points correspond to an efficiency of 60\% and 70\%, respectively. }
\label{fig:btageff}
\end{figure}

The combined uncertainty in the jet energy scale (JES) and resolution (JER) is estimated using untagged jets in $13~\TeV$ data and simulation by following the methods described in Ref.~\cite{PERF-2016-04}. The total uncertainty is found to be less than 2\% of the jet \pt~across the investigated mass range.

For $b$-tagged jets an additional uncertainty is assigned to the energy scale (bJES). It is estimated using MC samples and verified with data following the method described in Ref.~\cite{PERF-2012-01}. Firstly, the ratio of the sum of track transverse momenta inside the jet to the total jet transverse momentum measured in the calorimeter is formed, and then this ratio is compared between data and simulation. This double ratio is then compared for inclusive jets and $b$-jets. The relative uncertainty is found to be at most 2.6\% in the jet \pt~spectrum of interest and is applied in addition to the nominal jet energy scale uncertainty.

The uncertainties described above are summarized in Table \ref{tab:sys}. 
Other uncertainties that affect only the signal normalization, including the acceptance uncertainties associated with the choice of PDF and the uncertainty in the integrated luminosity, are found to be negligible.

\begin{table}[ht]
  \caption{Main sources of experimental systematic uncertainty in the low-mass and high-mass region. The background uncertainties quoted here are  the uncertainties on the fitted background yield in each mass bin. The JES, JER and bJES uncertainties quoted here are the uncertainties on the jet \pT. The $b$-jet trigger and $b$-tagging  uncertainties quoted here are the uncertainties on the per jet tagging efficiency.}
\label{tab:sys}
\begin{center}
\begin{tabular}{|l|c|c|c|}
\hline
  			                               &Low-mass region         & \multicolumn{2}{|c|}{High-mass region}\\
			                               &                                             &$\geq$1$b$ & 2$b$ \\\hline\hline
\multicolumn{4}{|c|}{			      Background Systematic Uncertainty }\\\hline
Fit function                                       &  (0.01 - 1.2)\%                           &  (0.01 - 9)\% & (0.01 - 10)\% \\
Fit parameters                                 & (0.2 - 2.7)\%                               &  (0.3 - 34)\% & (0.1 - 32)\%        \\\hline\hline
\multicolumn{4}{|c|}{			      Signal Systematic Uncertainty }\\\hline
JES                                                  & (1 - 1.2)\%                                 & \multicolumn{2}{|c|}{(1.2 - 1.6)\%}\\
JER                                                  &   (0.3 - 1)\%                               & \multicolumn{2}{|c|}{(0.3 -  0.4)\%}\\
bJES                                                &   2.6 \%                                   & \multicolumn{2}{|c|}{2.6 \%} \\
$b$-jet trigger                                  &   (1 - 20)\%                         & \multicolumn{2}{|c|}{$\cdots$} \\
$b$-tagging                                     &   (7 - 20 )\%                               & (3  - 5)\% & (10 - 30)\% \\\hline
\end{tabular}
\end{center}
\end{table}

\section{Interpretation}
\label{sec:result}

Since no significant deviation from the expected background is observed, limits are set on processes that would lead to resonances in the considered mass distributions.  The Bayesian method~\cite{BAT} is used to set 95\% credibility-level (C.L.) upper limits on the cross section, where the 95\% quantile of the posterior is taken as the upper limit.  A Gaussian prior is used for each nuisance parameter corresponding to a systematic uncertainty, and a flat prior is used for the signal normalization.  The expected limits as well as the $1\sigma$ and $2\sigma$ bands are calculated using pseudo-experiments generated from the background model by incorporating all systematic uncertainties in both the signal and background predictions.  Template morphing~\cite{Baak:2014fta} is utilized to interpolate between the resonance mass values of the signal hypotheses that are realized in MC simulation. The simulated models are described in detail in Sec.~\ref{sec:mc}. Theoretical uncertainties affecting the signal cross section are not considered.  Figs.~\ref{fig:Bstar}, \ref{fig:Zprime} and \ref{fig:DMZ} show the cross section limits for the $b^*$ signal using the inclusive $b$-jet selection, the $Z'$ signal using the combined low- and high-mass 2-$b$-tags selection, and the DM $Z'$ signal in the low- and high-mass 2-$b$-tags region, respectively.  The low- and high-mass selections overlap in the mass region between 1.2 and 1.5~\TeV.  For the combination the result with the better expected limit is chosen within the overlap region.

The cross section limits are translated to limits on the following signal mass parameters.  The $b^*$ model with an assumed branching fraction for $b^*\rightarrow bg$ of 85\% is excluded at 95\% C.L. for masses up to $2.6~\TeV$ using the inclusive single $b$-jet channel.  The double $b$-jet channel is used to set limits at 95\% C.L. which exclude masses up to $2.0~\TeV$ for the SSM $Z'\to b\overline{b}$ model and which exclude masses up to $2.1~\TeV$ for the leptophobic $Z'\to b\overline{b}$ model with SM-value couplings to quarks.  Mass limits on a dark-matter $Z'$ depend on the decay mode and the coupling strength to quarks, $g_{\text{SM}}$.  Assuming only the decay $Z'\to b\overline{b}$ and $g_{\text{SM}}=0.25$, masses up to $2.1~\TeV$ are excluded at 95\% C.L. Assuming $Z'$ decays to all five quark flavors other than the top quark and $g_{\text{SM}}=0.1$, masses up to $1.03~\TeV$ are excluded at 95\% C.L.

In order to allow for limit setting on new-physics models beyond those considered in the current studies, limits are quoted on the product of the cross section $\sigma$, acceptance $A$, selection efficiency $\epsilon$ and branching fraction $\mathcal{B}$ for a generic resonance with a reconstructed shape approximated by a Gaussian function, assuming a decay into two $b$-jets.    A MC-based transfer matrix is used to fold in the detector effects.  As the width is decreased from 15\% to 0\% of the mass, the cross section limits improve, but at the same time the limits are more affected by statistical fluctuations of the data in a single bin as compared to wider signals.  Fig.~\ref{fig:Gaus1} shows the limits for the inclusive $b$-jet selection when the intrinsic width is below the detector resolution. Fig.~\ref{fig:Gaus2} shows the corresponding limits for the low- and high-mass 2-$b$-tags selection.



\begin{figure}[!htb]
\begin{center}
\includegraphics[width=0.6\linewidth, angle=0]{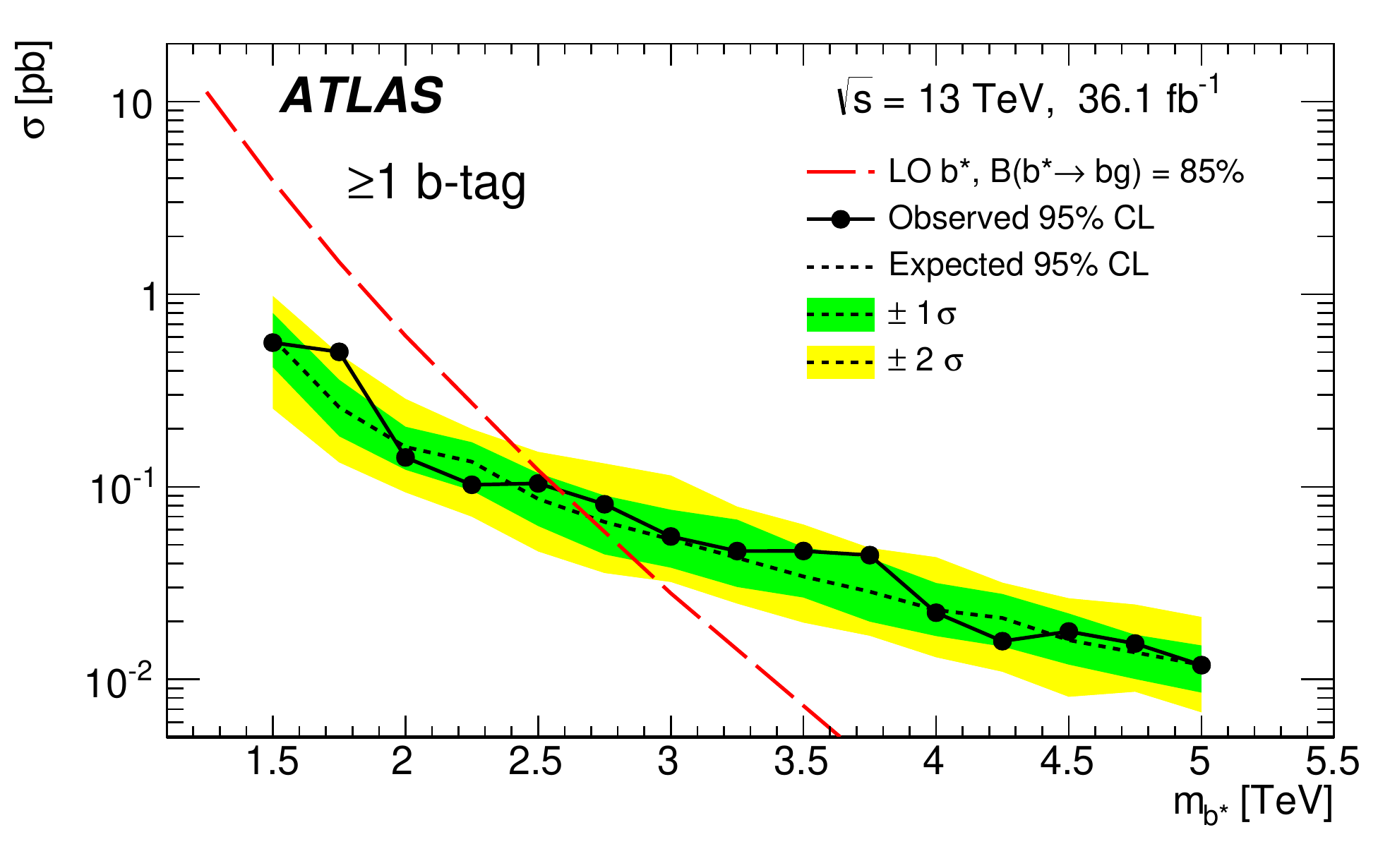}\caption{
Observed (filled circles) and expected (dotted line) 95\% credibility-level upper limits on the cross section for the $b^*$ model. The dashed lines show the predicted LO cross section as defined in Sec.~\ref{sec:mc}. The plot shows the results in the high-mass region with inclusive $b$-jet selection. 
\label{fig:Bstar}}
\end{center}
\end{figure}

\begin{figure}[!htb]
\begin{center}
\includegraphics[width=0.6\linewidth, angle=0]{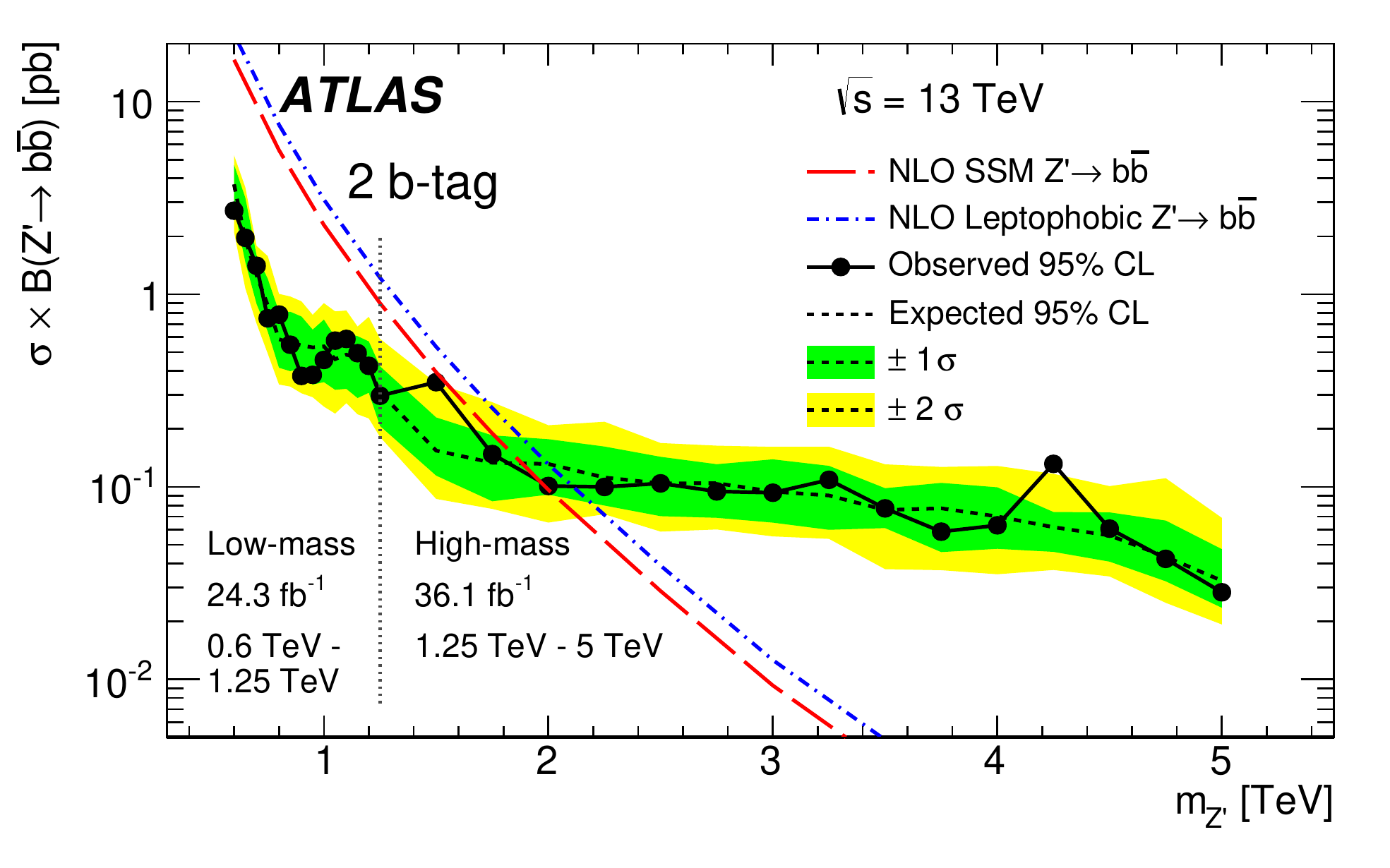}
\caption{Observed (filled circles) and expected (dotted line) 95\% credibility-level upper limits on the cross section times branching ratio for the SSM and leptophobic $Z'$ models. The dashed lines show the predicted NLO cross sections as defined in Sec.~\ref{sec:mc}. The plot shows the combined results in the low- and high-mass region (separated by the vertical dotted line) with two $b$-tags selection.
\label{fig:Zprime}}
\end{center}
\end{figure}

\begin{figure}
\begin{center}
    \subfigure[Low-mass two $b$-tags selection] {
\includegraphics[width=0.6\linewidth, angle=0]{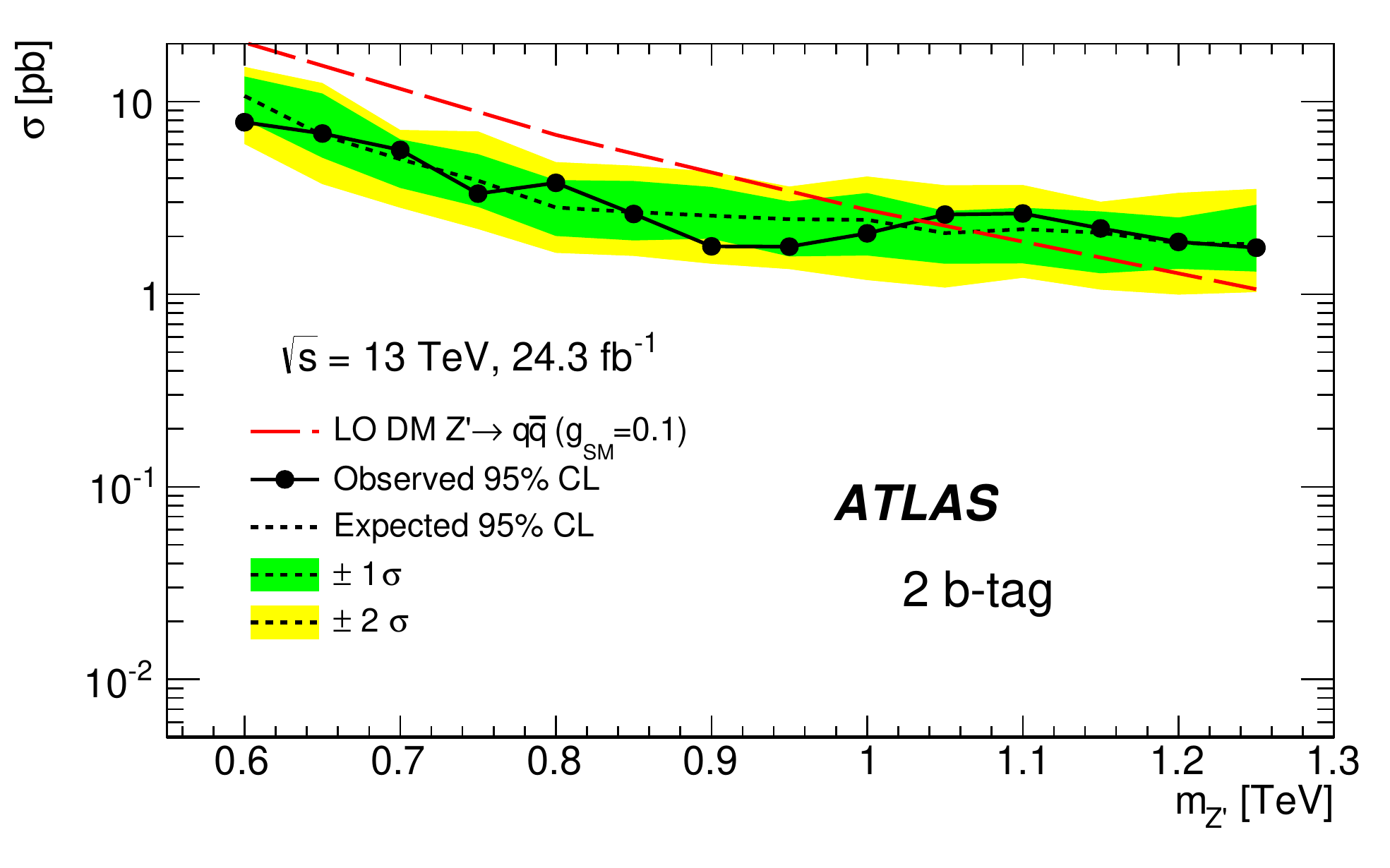}
}
    \subfigure[High-mass two $b$-tags selection] {
\includegraphics[width=0.6\linewidth, angle=0]{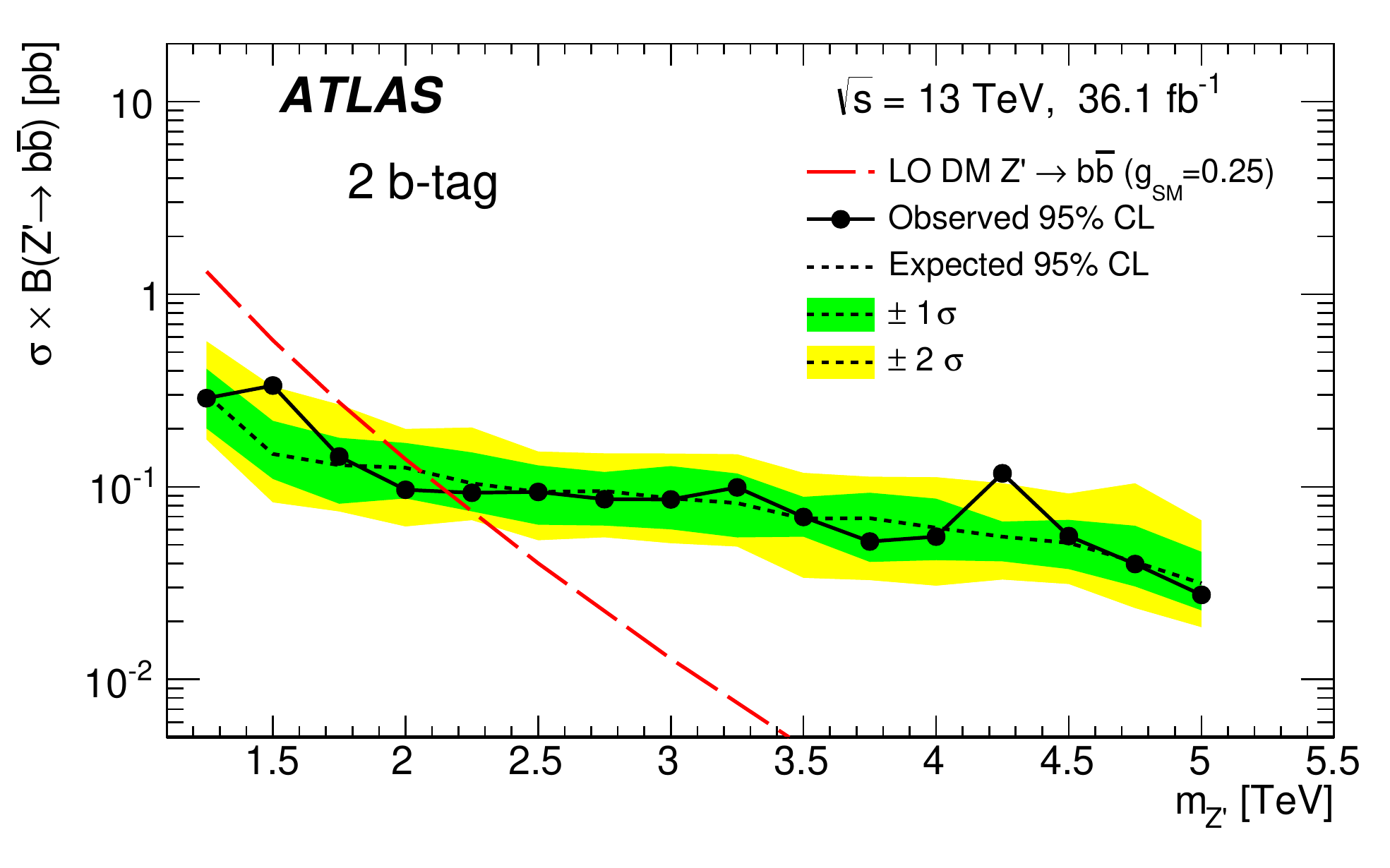}
}
\caption{Observed (filled circles) and expected (dotted line) 95\% credibility-level upper limits on the cross section for two different DM $Z'$ models.  In the low-mass region the $Z'$ is expected to decay to all five quark flavors other than the top quark and the mediator to SM quark coupling ($g_{\text{SM}}$) equal to 0.1 is assumed, whereas in the high-mass selection only the decays $Z'\to b\overline{b}$ are assumed with $g_{\text{SM}}=0.25$.  The dashed lines show the predicted LO cross sections as defined in Sec.~\ref{sec:mc}.
\label{fig:DMZ}}
\end{center}
\end{figure}

\begin{figure}[!htb]
\begin{center}
\includegraphics[width=0.6\linewidth, angle=0]{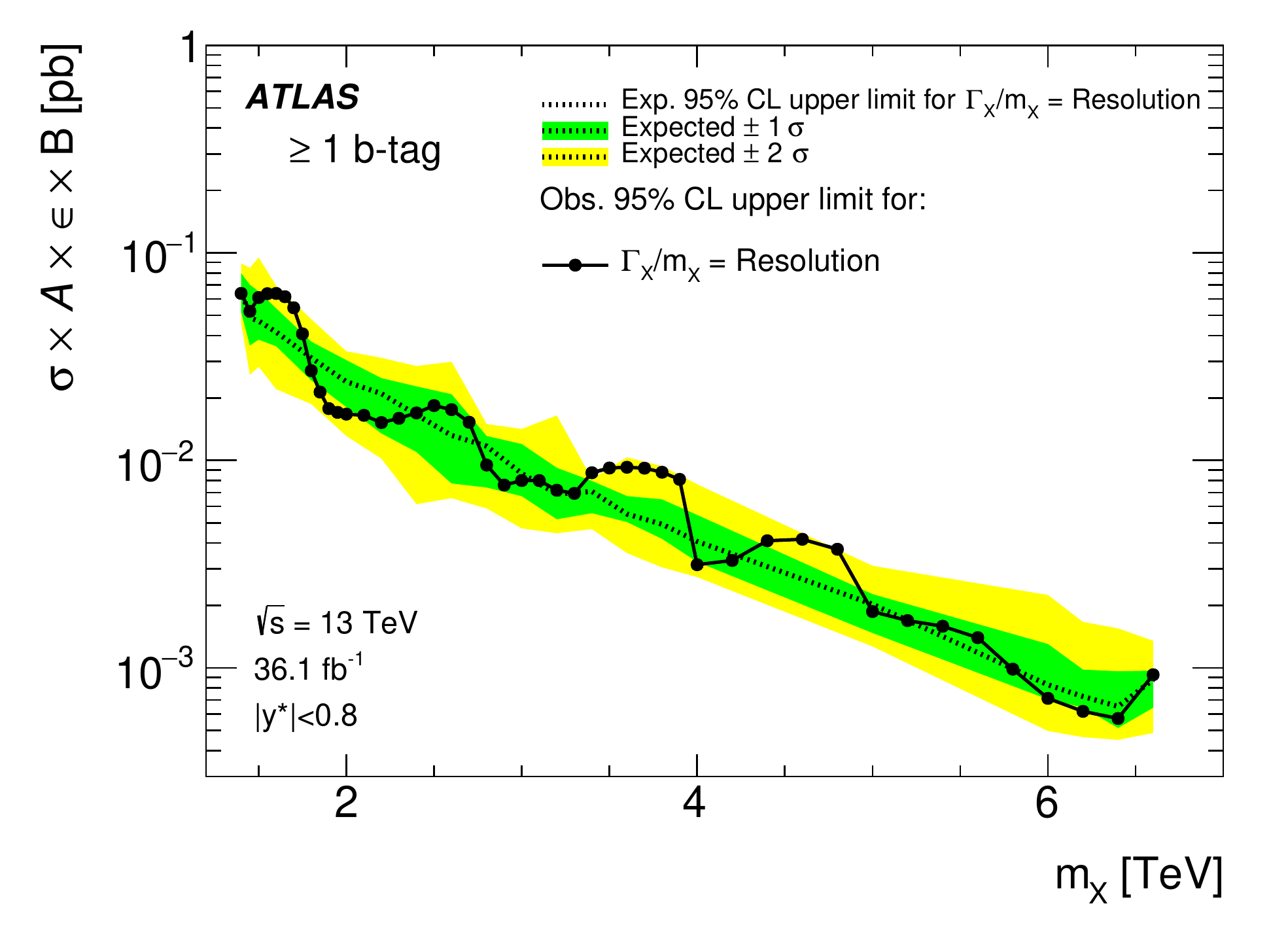}\caption{Observed (filled circles) and expected (dotted line)  95\% credibility-level upper limits on $\sigma \times A \times \epsilon \times {\mathcal{B}}(X \to b\bar{b})$, including kinematic acceptance and $b$-tagging efficiencies, for resonances with intrinsic width smaller than the detector resolution. The width of the Gaussian reconstructed shape is dominated by the dijet mass resolution. The plot shows the limits obtained from the high-mass inclusive $b$-jet selection.  
\label{fig:Gaus1}}
\end{center}
\end{figure}

\begin{figure}
\begin{center}
\includegraphics[width=0.6\linewidth, angle=0]{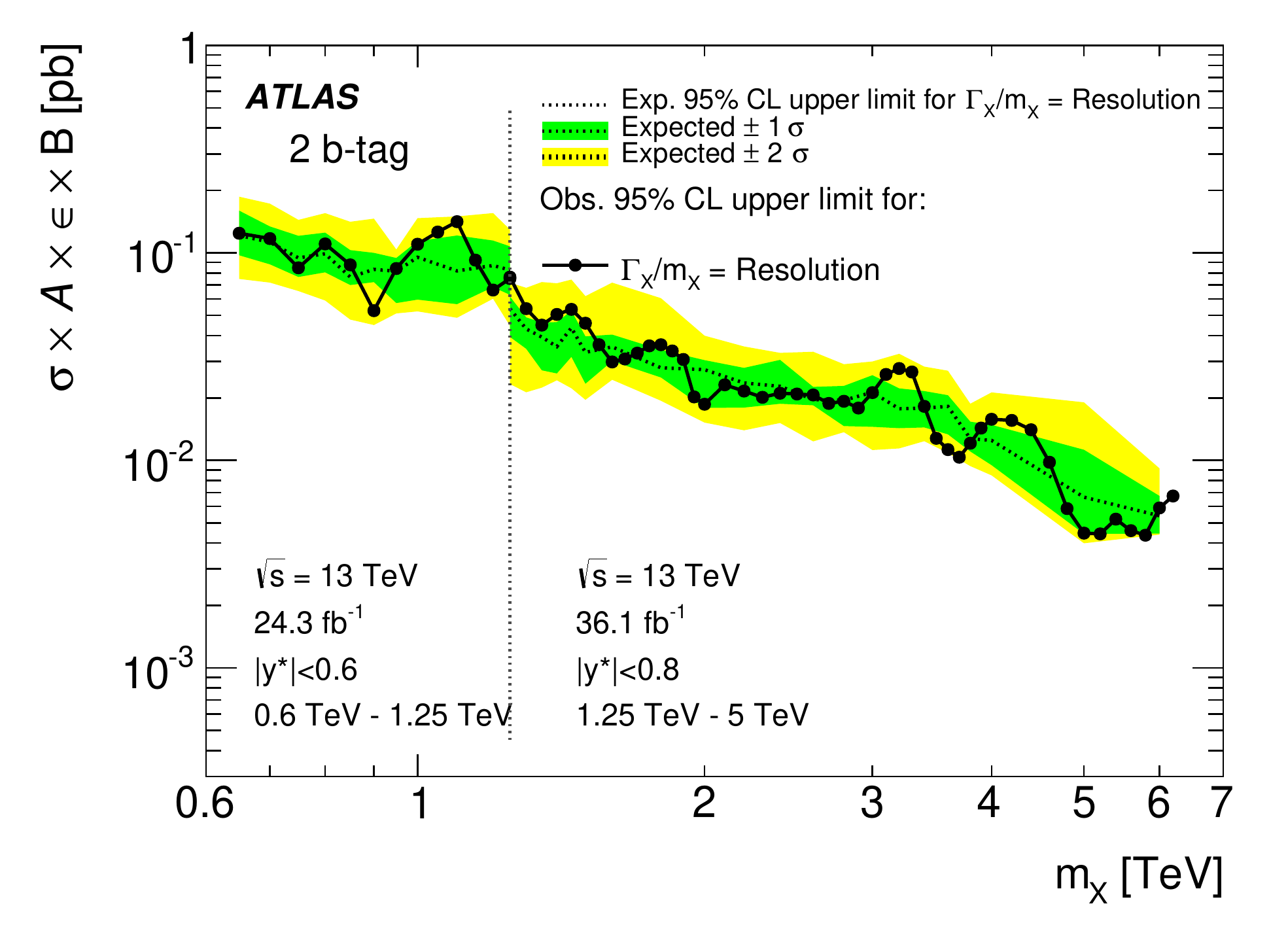}\caption{Observed (filled circles) and expected (dotted line)  95\% credibility-level upper limits on $\sigma \times A \times \epsilon \times {\mathcal{B}}(X \to b\bar{b})$, including kinematic acceptance and $b$-tagging efficiencies, for resonances with intrinsic width smaller than the detector resolution. The width of the Gaussian reconstructed shape is dominated by the dijet mass resolution. The plot shows the limits obtained from the combined low- and high-mass (separated by the vertical dotted line) two $b$-tags selection.  
\label{fig:Gaus2}}
\end{center}
\end{figure}

%
\FloatBarrier

\section{Conclusion}
\label{sec:conclusion}
Searches are performed for high-mass resonances in the dijet invariant mass spectrum with one or two jets identified as $b$-jets, using an integrated luminosity of up to $36.1~\ifb$ of proton-proton collisions with a center-of-mass  energy of $\sqrt{s}=13~\TeV$ recorded by the ATLAS detector at the Large Hadron Collider.
The search presented in this paper probes the mass range 0.57--5~\TeV.
No evidence of a significant excess of events above the expected Standard Model background is found.

Excited $b^*$-quarks with $b^* \to bg$ decays are excluded at 95\% C.L. for masses up to 2.6~\TeV. New $Z'$ gauge bosons are excluded in the sequential standard model (SSM) $Z'\to b\overline{b}$ model for masses up to 2.0~\TeV, and excluded in the leptophobic $Z'\to b\overline{b}$ model with SM-value couplings to quarks for masses up to 2.1~\TeV, both at 95\% C.L. Lastly, a $Z'$ axial-vector dark-matter mediator with only $b$-quark couplings set to $g_{\text{SM}}=0.25$ and axial DM couplings of $g_{\text{DM}}=1.0$, is excluded at 95\% C.L. for masses  up to 2.1~\TeV. Assuming $Z'$ decays into all five quark flavors other than the top quark and $g_{\text{SM}}=0.1$, masses up to $1.03~\TeV$ are excluded at 95\% C.L.

\section*{Acknowledgments}


We thank CERN for the very successful operation of the LHC, as well as the
support staff from our institutions without whom ATLAS could not be
operated efficiently.

We acknowledge the support of ANPCyT, Argentina; YerPhI, Armenia; ARC, Australia; BMWFW and FWF, Austria; ANAS, Azerbaijan; SSTC, Belarus; CNPq and FAPESP, Brazil; NSERC, NRC and CFI, Canada; CERN; CONICYT, Chile; CAS, MOST and NSFC, China; COLCIENCIAS, Colombia; MSMT CR, MPO CR and VSC CR, Czech Republic; DNRF and DNSRC, Denmark; IN2P3-CNRS, CEA-DRF/IRFU, France; SRNSFG, Georgia; BMBF, HGF, and MPG, Germany; GSRT, Greece; RGC, Hong Kong SAR, China; ISF, I-CORE and Benoziyo Center, Israel; INFN, Italy; MEXT and JSPS, Japan; CNRST, Morocco; NWO, Netherlands; RCN, Norway; MNiSW and NCN, Poland; FCT, Portugal; MNE/IFA, Romania; MES of Russia and NRC KI, Russian Federation; JINR; MESTD, Serbia; MSSR, Slovakia; ARRS and MIZ\v{S}, Slovenia; DST/NRF, South Africa; MINECO, Spain; SRC and Wallenberg Foundation, Sweden; SERI, SNSF and Cantons of Bern and Geneva, Switzerland; MOST, Taiwan; TAEK, Turkey; STFC, United Kingdom; DOE and NSF, United States of America. In addition, individual groups and members have received support from BCKDF, the Canada Council, CANARIE, CRC, Compute Canada, FQRNT, and the Ontario Innovation Trust, Canada; EPLANET, ERC, ERDF, FP7, Horizon 2020 and Marie Sk{\l}odowska-Curie Actions, European Union; Investissements d'Avenir Labex and Idex, ANR, R{\'e}gion Auvergne and Fondation Partager le Savoir, France; DFG and AvH Foundation, Germany; Herakleitos, Thales and Aristeia programmes co-financed by EU-ESF and the Greek NSRF; BSF, GIF and Minerva, Israel; BRF, Norway; CERCA Programme Generalitat de Catalunya, Generalitat Valenciana, Spain; the Royal Society and Leverhulme Trust, United Kingdom.

The crucial computing support from all WLCG partners is acknowledged gratefully, in particular from CERN, the ATLAS Tier-1 facilities at TRIUMF (Canada), NDGF (Denmark, Norway, Sweden), CC-IN2P3 (France), KIT/GridKA (Germany), INFN-CNAF (Italy), NL-T1 (Netherlands), PIC (Spain), ASGC (Taiwan), RAL (UK) and BNL (USA), the Tier-2 facilities worldwide and large non-WLCG resource providers. Major contributors of computing resources are listed in Ref.~\cite{ATL-GEN-PUB-2016-002}.



\printbibliography


\newpage

 
\begin{flushleft}
{\Large The ATLAS Collaboration}

\bigskip

M.~Aaboud$^\textrm{\scriptsize 34d}$,    
G.~Aad$^\textrm{\scriptsize 99}$,    
B.~Abbott$^\textrm{\scriptsize 124}$,    
O.~Abdinov$^\textrm{\scriptsize 13,*}$,    
B.~Abeloos$^\textrm{\scriptsize 128}$,    
D.K.~Abhayasinghe$^\textrm{\scriptsize 91}$,    
S.H.~Abidi$^\textrm{\scriptsize 164}$,    
O.S.~AbouZeid$^\textrm{\scriptsize 143}$,    
N.L.~Abraham$^\textrm{\scriptsize 153}$,    
H.~Abramowicz$^\textrm{\scriptsize 158}$
% ...
\bigskip
\\

% ...
$^{13}$Institute of Physics, Azerbaijan Academy of Sciences, Baku; Azerbaijan.\\
% ...
$^{34}$$^{(a)}$Facult\'e des Sciences Ain Chock, R\'eseau Universitaire de Physique des Hautes Energies - Universit\'e Hassan II, Casablanca;$^{(b)}$Centre National de l'Energie des Sciences Techniques Nucleaires (CNESTEN), Rabat;$^{(c)}$Facult\'e des Sciences Semlalia, Universit\'e Cadi Ayyad, LPHEA-Marrakech;$^{(d)}$Facult\'e des Sciences, Universit\'e Mohamed Premier and LPTPM, Oujda;$^{(e)}$Facult\'e des sciences, Universit\'e Mohammed V, Rabat; Morocco.\\
% ...
$^{91}$Department of Physics, Royal Holloway University of London, Egham; United Kingdom.\\
% ...
$^{99}$CPPM, Aix-Marseille Universit\'e, CNRS/IN2P3, Marseille; France.\\
% ...
$^{124}$Homer L. Dodge Department of Physics and Astronomy, University of Oklahoma, Norman OK; United States of America.\\
% ...
$^{128}$LAL, Universit\'e Paris-Sud, CNRS/IN2P3, Universit\'e Paris-Saclay, Orsay; France.\\
% ...
$^{143}$Santa Cruz Institute for Particle Physics, University of California Santa Cruz, Santa Cruz CA; United States of America.\\
% ...
$^{153}$Department of Physics and Astronomy, University of Sussex, Brighton; United Kingdom.\\
% ...
$^{158}$Raymond and Beverly Sackler School of Physics and Astronomy, Tel Aviv University, Tel Aviv; Israel.\\
% ...
$^{164}$Department of Physics, University of Toronto, Toronto ON; Canada.\\
% ...
$^{a}$ Also at  Department of Physics, University of Malaya, Kuala Lumpur; Malaysia.\\
$^{b}$ Also at Borough of Manhattan Community College, City University of New York, NY; United States of America.\\
$^{c}$ Also at Centre for High Performance Computing, CSIR Campus, Rosebank, Cape Town; South Africa.\\
$^{d}$ Also at CERN, Geneva; Switzerland.\\
$^{e}$ Also at CPPM, Aix-Marseille Universit\'e, CNRS/IN2P3, Marseille; France.\\
% ...
$^{*}$ Deceased

\end{flushleft}






\end{document}